\documentclass[floatfix, aps, prb, reprint]{revtex4-2}
\usepackage[T1]{fontenc}
\usepackage{bm}
\usepackage{graphicx}
\usepackage{amsmath}
\usepackage{amssymb}
\usepackage{color}
\usepackage{ifpdf}
\usepackage{chngcntr}
\usepackage{placeins}
\usepackage{enumitem}

\begin{document}
\title{\textbf{Band flattening in buckled monolayer graphene}}

\author{S. P. Milovanovi\'{c}}\email{slavisa.milovanovic@uantwerpen.be}
\author{M. An\dj elkovi\'{c}}\email{misa.andelkovic@uantwerpen.be}
\author{L. Covaci}\email{lucian@covaci.org}
\author{F. M. Peeters}\email{francois.peeters@uantwerpen.be}

\affiliation{Physics Department, University of Antwerp \\
Groenenborgerlaan 171, B-2020 Antwerpen, Belgium}
\begin{abstract}
The strain fields of periodically buckled graphene induce a periodic pseudo-magnetic field (PMF) that modifies the electronic band structure. From the geometry, amplitude, and period of the periodic pseudo-magnetic field, we determine the necessary conditions to access the regime of correlated phases by examining the band flattening. As compared to twisted bilayer graphene the proposed system has the advantages that: 1) only a single layer of graphene is needed, 2) one is not limited to hexagonal superlattices, and 3) narrower flat bandwidth and larger separation between flat bands can be induced. We, therefore, propose that periodically strained graphene single layers can become a platform for the exploration of exotic many-body phases.
\end{abstract}
\date{Antwerp, \today}
\maketitle
\section{Introduction}
Recently, twisted bilayers became a hot topic in 2D materials research due to the emergence of correlated phases when the rotation angle between the two layers is close to so called "magic angles" \cite{r_cao_1, r_cao_2, r_kim_1, r_sharpe, r_yank}. Theory predicted that by rotating two lattices the moir\'{e} modulation will influence the inter-layer electron tunneling within the moir\'{e} super-cell leading to the formation of moir\'{e} bands\cite{r_tfb_1, r_fb_bis, r_tfb_2, r_tfb_3, r_tfb_4, r_tfb_5}. The width of the bands depends on the rotation angle. However, this dependence is not monotonic. Bistritzer and MacDonald showed that at certain angles the band velocity drops to zero leading to extremely flat bands \cite{r_fb_bis}, i.e. "magic angles". This led to the discovery of superconducting \cite{r_cao_1} and Mott insulator \cite{r_cao_2} phases in twisted bilayer graphene rotated by approximately 1.1$^\circ$.

However, adjusting the angle between the two monolayers is nothing but a challenging task. The process is slow and lacks sufficient control over the rotation angle. On the other hand, flat bands are not a phenomenon solely related to twisted bilayer graphene. Recently, researchers found that in the case of ABC-stacked trilayer graphene one can create flat bands just by placing a hBN layer on top of it \cite{r_tri_1, r_tri_2, r_tri_3}. The already flat, low energy cubic bands of the trilayer become significantly narrower due to a moir\'{e} induced modulation even without twisting two materials. Furthermore, it was shown that the bandwidth can be controlled by electrostatic gating. 

In previous examples, because of the hexagonal graphene and hBN lattices, only hexagonal moir\'{e} lattice structures can be realized. This moir\'{e} modulation had a key role in the formation of flat bands. But this is not the only way to create flat bands. In fact, any periodic modulation that is strong enough to localize electrons will result in modulated super-lattice bands whose width will depend on parameters of that modulation. For example, a periodic magnetic field is predicted to result in flat bands \cite{r_rmf_0, r_rmf_1, r_rmf_2, r_rmf_3}, where the bandwidth depends both on the period of the field and its magnitude. The advantage of such an external induced periodic magnetic field is that one is not restricted to the symmetries of the lattice of the material like in the case of twisted structures. This provides additional degrees of freedom to manipulate electron densities and control electron correlations. However, it comes with a price. Usually, extremely strong fields of order of tens (or hundreds) of Tesla are needed to sufficiently reduce the band width to make it smaller than the strength of the inter-particle interactions. 

More promising approaches are based on the use of pseudo-magnetic fields (PMF) rather than real fields. In a recent study, Jiang \textit{et al.} reported the emergence of flat bands and a Mott insulating phase in a periodically buckled graphene lattice \cite{r_fb_01}. The periodically buckled graphene can appear because of different reasons: due to different thermal expansion coefficients of the graphene and its substrate\cite{r_ann_1}, during the growth due to defects in the substrate \cite{r_def_1}, doping of the substrate \cite{r_dop_1}, straining its lattice \cite{r_fb_03}, or even due to strains in twisted structures \cite{r_fb_02}. Zhang \textit{et al.} showed theoretically that 2D buckling modes appear in case of graphene on hBN in the presence of strong biaxial strain \cite{r_bu_mo1}. The induced potential acts as a pseudo-magnetic field on electron motion, which provides additional knobs to tune the bandwidth. Unlike real magnetic fields, strain induced pseudo-magnetic fields have been shown to generate strong fields of few hundreds of Tesla \cite{r_crommie}. Hence, the fast development of the field and the lack of theoretical studies motivated us to perform an extensive study of the appearance of flat bands in periodically strained structures.

This paper is organized as follows. In Sec. \ref{sec::1} we introduce the theoretical model and present the numerical methods that are used to simulate buckled systems. In Sec. \ref{sec::2} we give the results for the triangular PMF mode. Results for the hexagonal and herringbone buckling modes are given in Sec. \ref{sec::3}. We calculate the spatial LDOS maps as well as current plots for the three low energy flat bands in the case of triangular PMF and the hexagonal buckling modes. These results are shown in Sec. \ref{sec::4}. In Sec. \ref{sec::4a} we study the influence of a homogeneous real magnetic field on flat bands. Lastly, our final remarks and conclusions are given in Sec. \ref{sec::5}.
\section{Methods}
\label{sec::1}
To calculate the electronic properties of buckled two-dimensional sheets we use the nearest-neighbour tight-binding (TB) model given by
\begin{equation}
    \label{e_tb1}
    H = \sum_{i,j} t_{ij}(\mathbf{r})c_{i}^{\dagger}c_j,
\end{equation}
where $c_{i}^{\dagger}$($c_i$) is the creation (annihilation) operator for an electron at site $i$ and $t_{ij}$ is the hopping energy between sites $i$ and $j$. The numerical results presented in this paper are calculated using two user-friendly software packages: Pybinding \cite{r_pb} and KITE \cite{r_kite}, designed for fast generation of tight-binding Hamiltonians and supported with efficient solvers for calculations of different electronic properties of TB systems. Notice that $t_{ij}$ in Eq. \eqref{e_tb1} is a spatially dependent function. The spatial distribution of the hopping integral is solely determined by the straining configuration which defines the profile of the pseudo-magnetic field. The change of the equilibrium positions of atoms in a graphene sheet by strain is mirrored in the change of the hopping energies as
\begin{equation}
\label{e_tb02}
t_{ij}(\mathbf{r}) = t_0 e^{-\beta(r_{ij}/a_0 - 1)},
\end{equation}
where $t_0 = -2.8$ eV is the equilibrium hopping energy, $a_0=0.142$ nm is the length of the unstrained C-C bond, and $r_{ij} = \left| \mathbf{r}_i - \mathbf{r}_j\right|$ is the length of the strained bond between atoms $i$ and $j$. The decay factor $\beta = \partial \log t / \partial \log a \mid_{a=a_0} \approx 3.37$ describes the change of the hopping energy with the modulation of the bond length \cite{r_tb_1}. 

The spatial variation of the hopping energy is equivalent to the generation of a pseudo-magnetic vector potential, $\mathbf{A} = (A_x, A_y, 0)$, which can be evaluated around the $\mathbf{K}$ point using\cite{r_tb_1}
\begin{equation}
\label{e_vecpot}
A_x - \mathtt{i} A_y = -\frac{1}{ev_F}\sum_j \delta t_{ij} e^{\mathtt{i}\mathbf{K}\cdot \mathbf{r_{ij}}},
\end{equation}
where the sum runs over all neighboring atoms of atom $i$, $v_F = 3ta_{0}/2$ is the Fermi velocity, and $\delta t_{ij} = (t_{ij} - t_0)$.
It is known that by using a linear expansion of Eq. \eqref{e_vecpot} one can easily connect the vector potential to the strain tensor $\pmb{\varepsilon}$ in the following way \cite{r_linapp}
\begin{equation}
\label{ch8e23}
\mathbf{A} = A_0 \begin{pmatrix}
\varepsilon_{xx} - \varepsilon_{yy} \\ -2\varepsilon_{xy}
\end{pmatrix},
\end{equation}
where $\varepsilon_{ij}$ are the elements of the strain tensor and $A_0=-\frac{\hbar \beta}{2ea_{cc}}$. Such a linear expansion has been shown to be a good approximation up to strain of 10 $\%$ \cite{r_dean_1}.
The pseudo-magnetic field is then obtained as
\begin{equation}
\label{ch8e3}
\mathbf{B_{ps}} = \pmb{\bigtriangledown} \times \mathbf{A}.
\end{equation}

Here, we will consider three different geometries of PMFs generated by three different buckling profiles: \textit{triangular PMF mode}, \textit{hexagonal buckling mode}, and \textit{herringbone buckling mode}.

The triangular PMF mode has been experimentally observed in Ref. \onlinecite{r_fb_01} where the authors reported on the existence of flat bands due to periodic PMF generated by the two-dimensional buckling in graphene and, therefore, we will study this configuration in detail. The hexagonal buckling mode is chosen since the out-of-plane deformation of the hexagonal mono-layers stacked on top of each other (e.g. graphene on hBN, twisted-bilayer graphene, etc.) has hexagonal symmetry \cite{r_oop_1, r_oop_2, r_oop_3}, while the herringbone mode is shown to be the lowest-energy solution of all buckling modes in the case of large straining \cite{r_bu_mo1} and consequently, of high interest for buckled structures.

The deformation fields given by these different modes are plugged into the Hamiltonian by changing the hopping energy using Eq. \eqref{e_tb02}. In the case of a triangular PMF profile the deformation fields are unknown. Hence, we need to reverse the problem for this case and starting from the pseudo-magnetic field obtain the vector potential. The procedure is similar to the one given in Ref. \onlinecite{r_rev_1} and the details can be found in Appendix \ref{sec::s1}. Due to the gauge invariance, the solution for the deformation field is not uniquely defined. Nonetheless, the choice of the gauge shouldn't be of importance in the low-energy spectrum where the scattering mechanism is defined by the PMF. This is unlike the high-energy regime, e.g. beyond 1 eV, where the gauge choice defines the position of the van Hove singularities. 
\section{Triangular PMF mode}
\label{sec::2}
\begin{figure*}[bt]
\begin{center}
\includegraphics[width=0.9\textwidth]{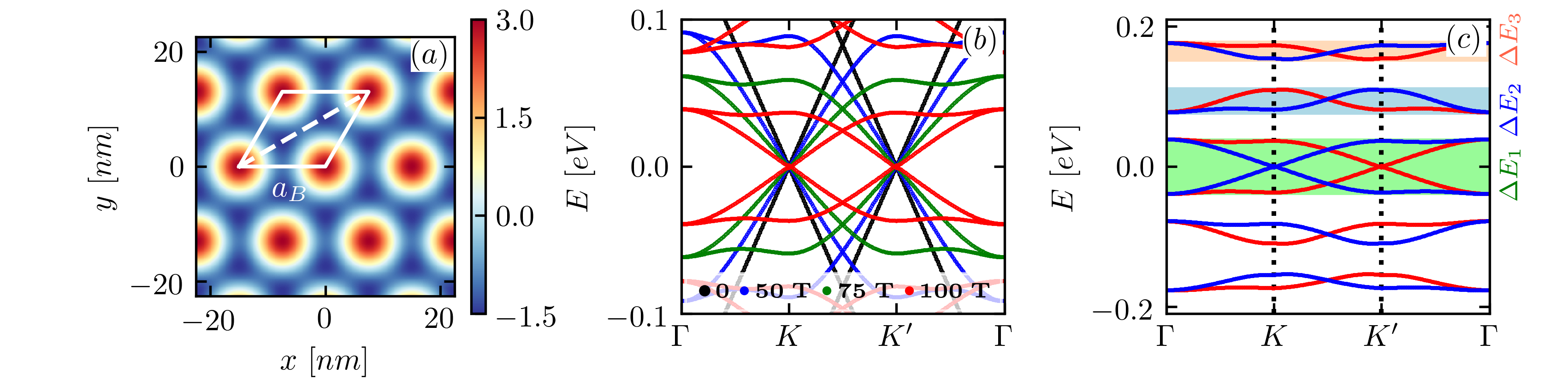}
\caption{(a) Profile of the pseudo-magnetic field given by Eq. \eqref{e_tpmf_1} with a unit cell shown by the white rhombus with $a_B = 15$ nm. The field is given in units of $B_0$. (b) Corresponding band structure for a few different values of $B_0$ shown in the inset. (c) Band structure, for $B_0 = 100$ T, showing three low energy mini-bands marked by green, blue and orange rectangles. Blue and red color curves show bands belonging to different valleys of graphene.}
\label{fig_1}
\end{center}
\end{figure*}

We start by showing in Fig. \ref{fig_1}(a) the profile of the triangular PMF mode. It is given by
\begin{equation}
\label{e_tpmf_1}
B(\mathbf{r}) = B_0 \sum_{i = 1}^3 \cos(\mathbf{K}_i \mathbf{r}), 
\end{equation}
where $B_0$ is the amplitude of the field, $\mathbf{K}_1 = K\frac{2}{\sqrt{3}}\vec{e}_y$, $\mathbf{K}_2 = K(\vec{e}_x + \frac{1}{\sqrt{3}}\vec{e}_y)$, $\mathbf{K}_3 = \mathbf{K}_1 + \mathbf{K}_2$, and $K = 2\pi/a_B$, with $a_B$ being the period of the PMF unit cell which is shown by the white rhombus in Fig. \ref{fig_1}(a). In real space, unit vectors of the PMF unit cell are given by $\mathbf{a_{1, 2}} = n \mathbf{a_{1,2}^{gr}}$ where $\mathbf{a_{1}^{gr}} = (\sqrt{3}a_0, 0)$ and $\mathbf{a_{2}^{gr}} = (a_0/2, \sqrt{3}a_0/2)$ are the unit vectors of graphene and $n$ is an integer that satisfies $n\left|\mathbf{a_{1,2}^{gr}}\right| = a_B$. Consequently, in our calculations we use discrete values of $a_B$ that correspond to multiple integers of the length of the unit cell of graphene. Notice that the average field is zero and that the maximum (minimum) of the periodic field is $3B_0$ ($-1.5B_0$). The band structure of this system is given in Fig. \ref{fig_1}(b) for the case of no buckling (black curve) and for $B_0 = 50$ T (blue curve), 75 T (green curve), and 100 T (red curve) using $a_B = 15$ nm. One can see that the slope of graphene's linear bands decreases as the amplitude of the field is increased. The mechanism is the same as in the case of real magnetic field or by the induced potential in case of twisted bilayers, i.e. the periodic potential (in this case imposed by strain) modifies the linear graphene band structure into a series of mini-bands separated by relatively large gaps between adjacent sub-bands. In contrast to the twisted bilayer case, extremely flat bands do not occur at a discrete set of angles but rather, the band width continuously decreases with the strength of the amplitude of the field, as shown by the different curves in Fig. \ref{fig_1}(b). Since gaps appear, it is also interesting to know what happens with the widths of higher energy bands. In Fig. \ref{fig_1}(c) we show three lowest energy bands for $B_0 = 100$ T and $a_B = 15$ nm. Higher bands are much narrower than the lowest band, hence, of potential interest for the physics of correlated states. Furthermore, the pseudo-magnetic field breaks the valley symmetry of the system and lifts the degeneracy between the states belonging to different valleys (red and blue curves). In the following, we will examine these three bands in more detail.
\begin{figure*}[htbp]
\begin{center}
\includegraphics[width=14cm]{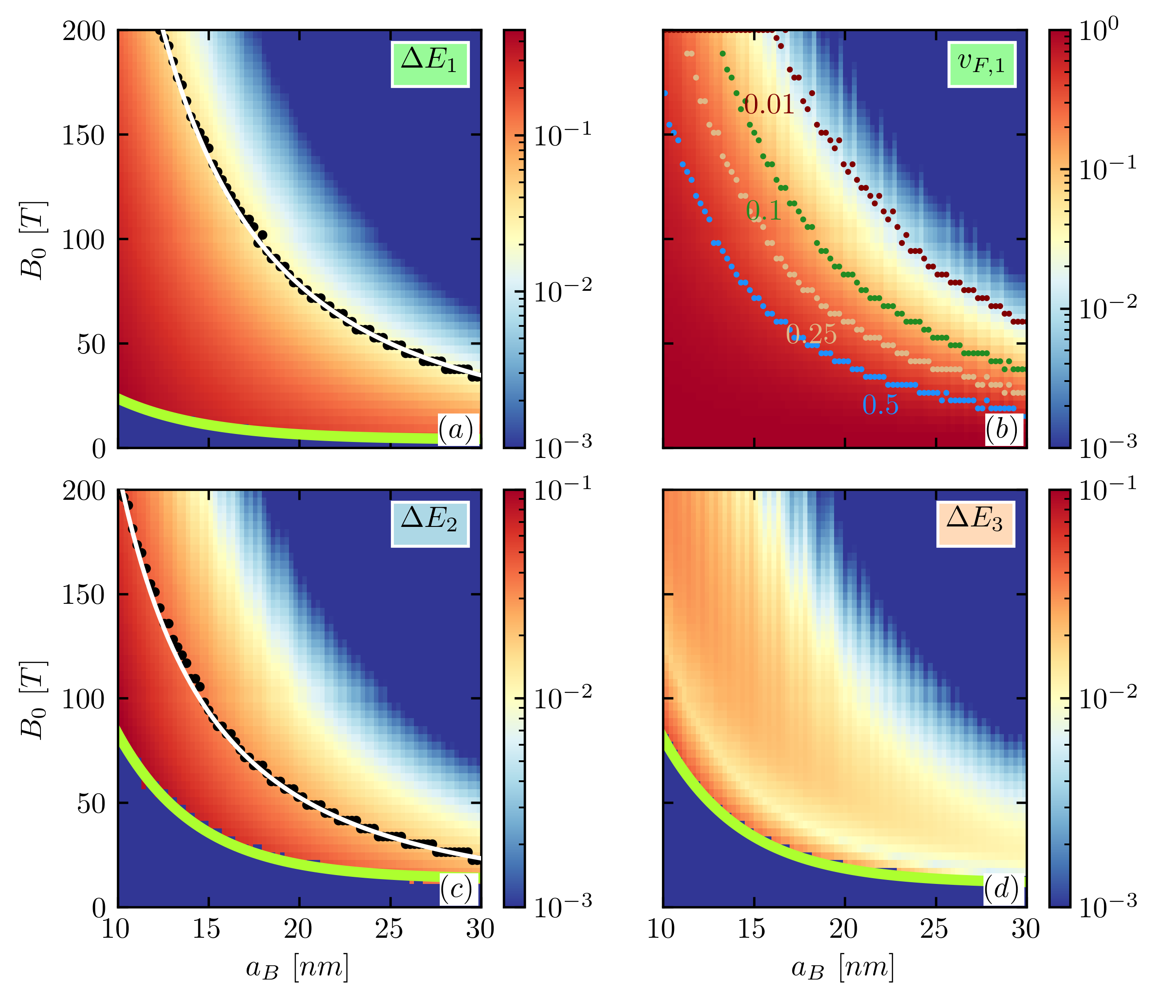}
\caption{(a) Contour plot of the width of the lowest energy band, $\Delta E_1$, (in eV) versus the length of the magnetic unit cell, $a_B$, and strength of the PMF. Black dots show values of $B_0$ at which the band width becomes smaller than $E_I$ and the white curve shows corresponding numerical fit. The green curve shows the value of the magnetic field at which the gap opens between the sub-bands. (b) Reduced Fermi velocity versus $a_B$ and $B_0$ of the lowest energy band at the Dirac point. The blue, orange, green, and maroon curve show the values of $B_0$ at which the Fermi velocity is reduced to 50 $\%$, 25 $\%$, 10 $\%$, and 1 $\%$ of its original value, respectively. (c) The same as (a) but for the second band. (d) The same as (a) but for the third band.}
\label{fig_2}
\end{center}
\end{figure*}

The width of the lowest energy band versus the period of the buckling, $a_B$, and the amplitude of the pseudo-magnetic field is shown in Fig. \ref{fig_2}(a). Notice that band flattening occurs with increasing magnitude of the magnetic field (for constant $a_B$) or by increasing the period of buckling (for constant $B_0$). For small values of the pseudo-magnetic field the spectrum is continuous. In Fig. \ref{fig_2}(a), this regime is located below the green curve, however, for plotting purposes we set the value of the band width here to be zero (the same for plots in Figs. \ref{fig_2}(c-d)). The value of the field at which the band gap opens between the lowest and the next band (green curve) can be fitted by following formula
\begin{equation}
\label{e_gap_1}
    B_1^{gap} = 0.4\frac{\Phi_0}{S},
\end{equation}
where $\Phi_0$ is the magnetic flux quantum and $S = \sqrt{3}a_B^2/2$ is the area of the unit cell. On the other hand, in order to have significant electron interactions in flat bands their width has to be smaller (or comparable to) the characteristic energy scale for the interactions \cite{r_kim_1} given by $E_I = e^2/4 \pi \varepsilon_0 \varepsilon a_B$. In Fig. \ref{fig_2}(a) we marked by black dots values of $B_0$ at which the band width of the lowest band is equal to this value (here we use $\varepsilon = 3$ which corresponds to graphene on hBN). The plot shows fast decay with the period of the unit cell and the dependence can be fitted by
\begin{equation}
\label{e_int_1}
    B_1^{int} = 6.5\frac{\Phi_0}{S},
\end{equation}
as shown by the white curve. The band flattening is followed by a decrease of the Fermi velocity, which is shown in Fig. \ref{fig_2}(b). Here, we plot the Fermi velocity calculated around the $\mathbf{K}$ point and scaled with its unstrained value. The curves of different colors show the values of the PMF for which the Fermi velocity reaches $50 \%$, $25 \%$, $10 \%$, and $1 \%$ of its original value.

We apply a similar approach for the second and the third mini-band from Fig. \ref{fig_1}(c). We find that the corresponding values of the critical field for the band-gap opening ($B_n^{gap}$) and the field at which the band width becomes smaller that the interaction energy ($B_n^{int}$) can be also fitted by a function of type $B_n^x = c_n^x \Phi_0/S$. In the case of the second band, the band gap opening occurs at $c_2^{gap} = 1.7$ (green curve in Fig. \ref{fig_2}(c)). At these fields, the band gap between third and the higher bands has already appeared, thus, we will use this value as the reference one. The second band becomes narrow enough for the electron interactions to be of significance at $c_2^{int} = 4.4$ (white curve in Fig. \ref{fig_2}(c)), which is well below the value for the lowest band. In the case of the third band (see Fig. \ref{fig_2}(d)), we find that as soon as the band gap appears (at $1.7\Phi_0/S$), its width is already below this criterion. 
\begin{figure*}[htbp]
\begin{center}
\includegraphics[width=17.5cm]{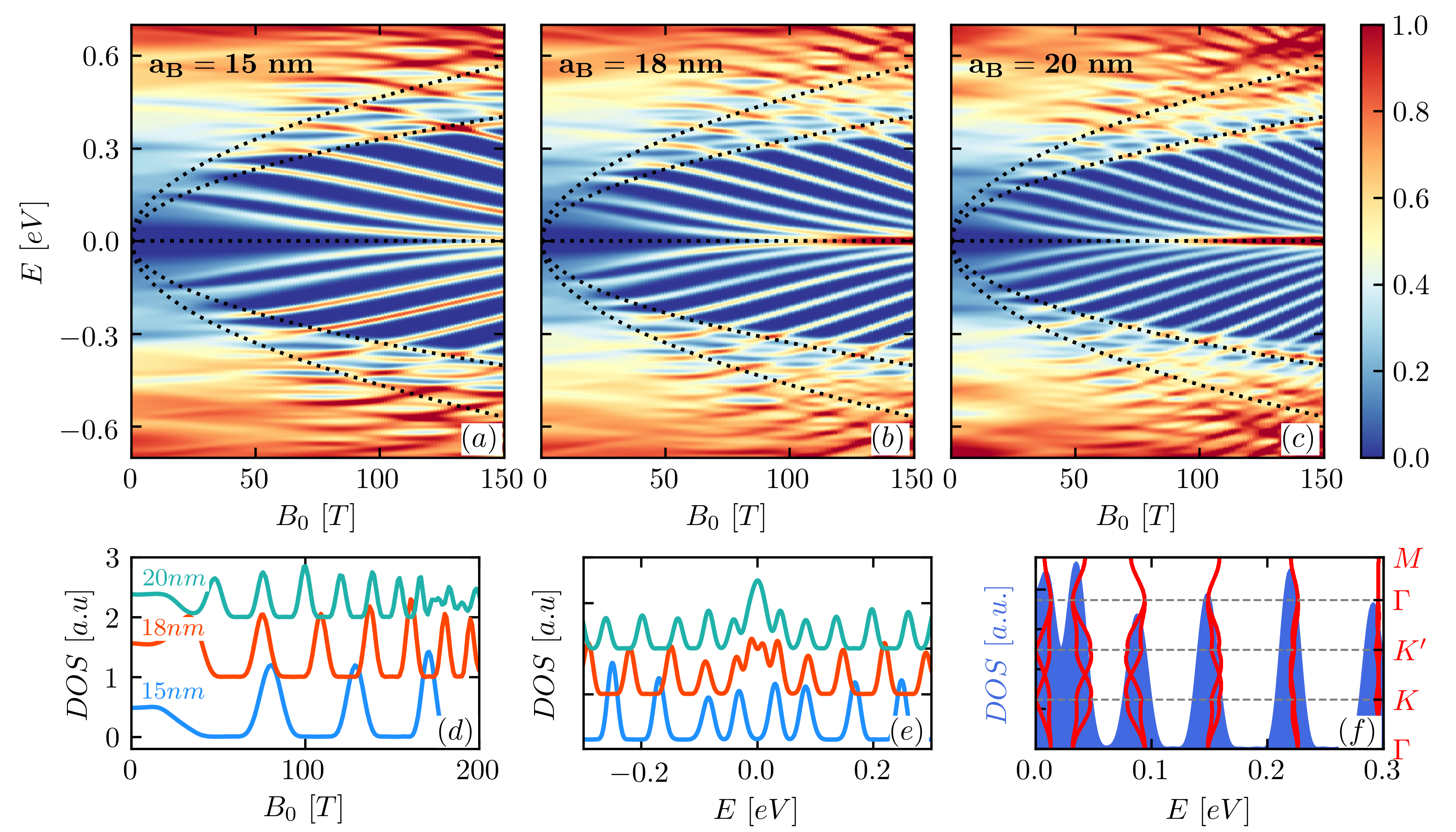}
\caption{Density of states (DOS) versus the amplitude of the PMF and energy for a unit cell with period (a) $a_B = 15$ nm, (c) $a_B = 18$ nm, and (a) $a_B = 20$ nm. Black dotted lines show Landau levels obtained using the analytic expression. (d) Cuts of the DOS at $E = 0.2$ eV for three different values of $a_B$ shown in (a-c). (e) Cuts of the DOS at $B_0 = 100$ T for three different values of $a_B$ shown in (a-c). (f) DOS (blue curve) and the band structure (red curves) calculated using $a_B = 18$ nm and $B_0 = 100$ T. DOS is calculated using energy broadening of 5 meV.}
\label{fig_3}
\end{center}
\end{figure*}

To further investigate effects of the periodic strain on the electronic properties of graphene we numerically calculate the density of states (DOS) versus the magnitude of the field and energy for three different periods of the PMF and the results are shown in Figs. \ref{fig_3}(a-c) for $a_B = 15$, $18$, and $20$ nm, respectively. These values are chosen since they agree well with the periods reported in Ref. \onlinecite{r_fb_01}.  In all three plots, one can spot series of well defined levels in the low energy spectrum. These levels are different from Landau levels (LL) which are given by $E_N = \pm v_f \sqrt{2e\hbar N B_{0}}$, where N is the LL index. The three lowest Landau levels, calculated using this analytic formula, are shown in the plots by black dotted lines. The zeroth and first LL are clearly visible in these plots while the trace of the second LL is only visible at very high fields and for larger periods of the superlattice. Notice that as the period of the unit cell is increased, the zeroth LL develops at lower field. The reason for this is the fact that the PMF varies fast over the unit cell and, consequently, Landau levels cannot develop. In order for Landau levels to appear, one needs to have a fairly constant field on the length scale of a few magnetic lengths \cite{r_llarea_0, r_llarea_1}.
\begin{figure*}[htbp]
\begin{center}
\includegraphics[width=15cm]{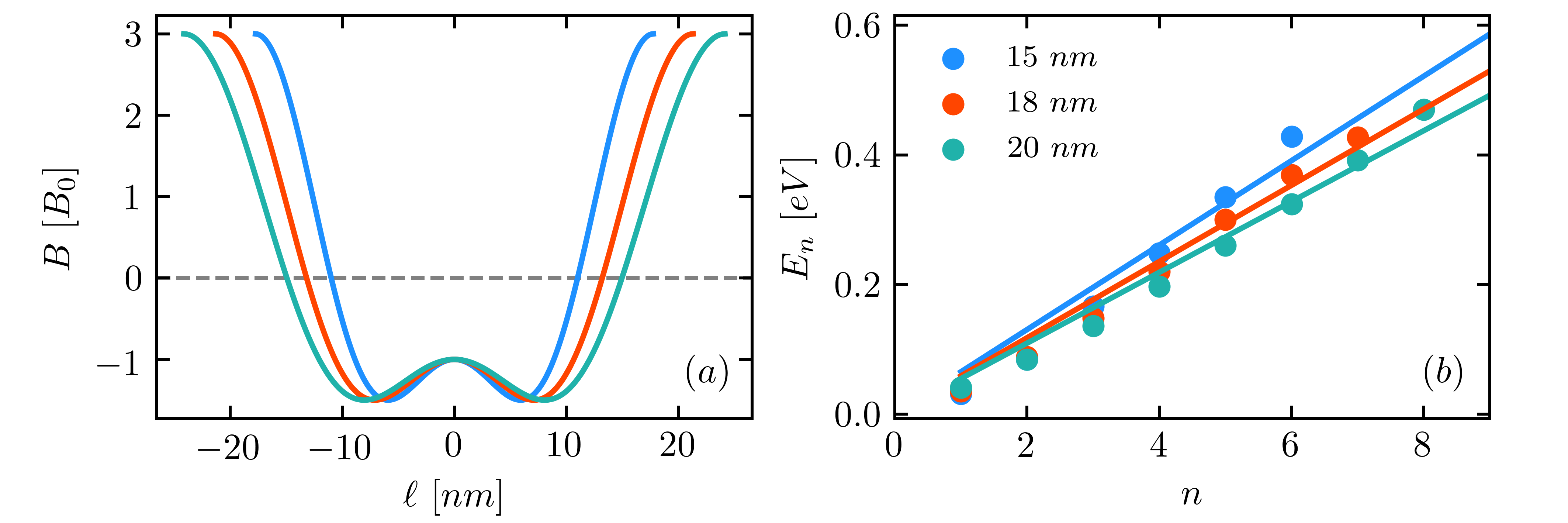}
\caption{(a) The profile of the pseudo-magnetic field along the path shown in Fig. \ref{fig_1}(a) by the dashed white line for three different values of $a_B$ given in (b). Horizontal dashed line shows $B = 0$ cut. (b) The energy position of the peaks in the DOS for $B_0 = 100$ T versus peak index for three different values of $a_B$ shown in the inset. Solid lines show corresponding linear fit to $E_n = n \Delta E$ with $\Delta E = $ 84, 73, and 66 meV for $a_B = $ 15, 18, and 20 nm, respectively.}
\label{fig_3a}
\end{center}
\end{figure*}

Unlike the Landau levels, well defined energy levels exist between them and decrease with the increase of the magnetic field until they eventually merge with the zeroth LL. This behaviour resembles the field-dependence of eigenstates of a graphene quantum dot \cite{r_qd_1} in an external magnetic field. Cuts of the DOS at a constant energy of $E = 0.2$ eV are shown in Fig. \ref{fig_3}(d) and constant magnitude of the PMF of $B_0 =  100$ T in Fig. \ref{fig_3}(e). In both cases, an increase of the period results in a decrease in the separation between the peaks. To strengthen our analogy with quantum dots, we point out that Libisch \textit{et al.} reported in Ref. \onlinecite{r_qd_2} the appearance of equidistant peaks in the DOS of a graphene quantum dot. The peaks are the consequence of the quantum well confinement in a system with a linear energy spectrum which results in equidistant energy levels with separation given by 
\begin{equation}
\label{e_es_1}
   \Delta E = \frac{\hbar v_F \pi}{W},
\end{equation}
where $v_F$ is the Fermi velocity and $W$ is the characteristic confinement length. The analogy between quantum dots and our system is justified since the strong periodic straining can be used to confine electrons, e.g. experiments on quantum emission in 2D materials on pillars \cite{r_np00, r_np01, r_np03, r_np06}. The idea is that the spatially varying vector potential generated by Eq. \eqref{ch8e23} modifies locally the electronic properties of the material and creates traps for electrons. In other words, strain is used as an instrument to form and control localized states in the system. The principle is illustrated in Fig. \ref{fig_3a}(a) where we show cuts of the pseudo-magnetic field along the dashed line in Fig. \ref{fig_1}(a) which form a PMF-defined well that acts as a trap for electrons. In our case, spatially varying field has a profound influence on the electronic behavior as electrons try to avoid energetically unfavorable regions of high PMF amplitude. The width of the well at $B = 0$ (dashed line in \ref{fig_3a}(a)) is 22, 25, and 29 nm for $a_B =$ 15, 18, and 20 nm, respectively. The peaks that we observe in Fig. \ref{fig_3}(e) are almost equidistant as Eq. \eqref{e_es_1} predicts (except in vicinity of Landau levels) and the dependence on the band index is fairly linear, as shown in Fig. \ref{fig_3a}(b). The small deviations from linearity are due to the fact that the boundaries of our confined region are not rigid since they are defined by the induced potential which depends on $B_0$ and the energy. Applying the previous formula to the average value of the peak separation from Fig. \ref{fig_3}(e) we obtain $W \approx 22$, 26, and 30 nm for $a_B =$ 15, 18, and 20 nm. This compares to the width of the PMF well at $B = 0$, shown in Fig. \ref{fig_3a}(a).

Having in mind our previous discussion and the results shown in Fig. \ref{fig_2} one can infer that the peaks in the DOS correspond to the flat bands as is evident from Fig. \ref{fig_3}(f) where we plot DOS for $B_0 = 100$ T and $a_B = 18$ nm together with the band structure for the same system. The two plots match perfectly. Thus, the levels in Figs. \ref{fig_3}(a-c) show the evolution of flat bands with the magnetic field. It is noteworthy to stress that these quasi-localized states forming the flat bands are not Landau levels. As will be shown later, these are scattering states, scattered by the strong PMF regions, and localized in the low magnetic field regions.
\begin{figure*}[htbp]
\begin{center}
\includegraphics[width=17.5cm]{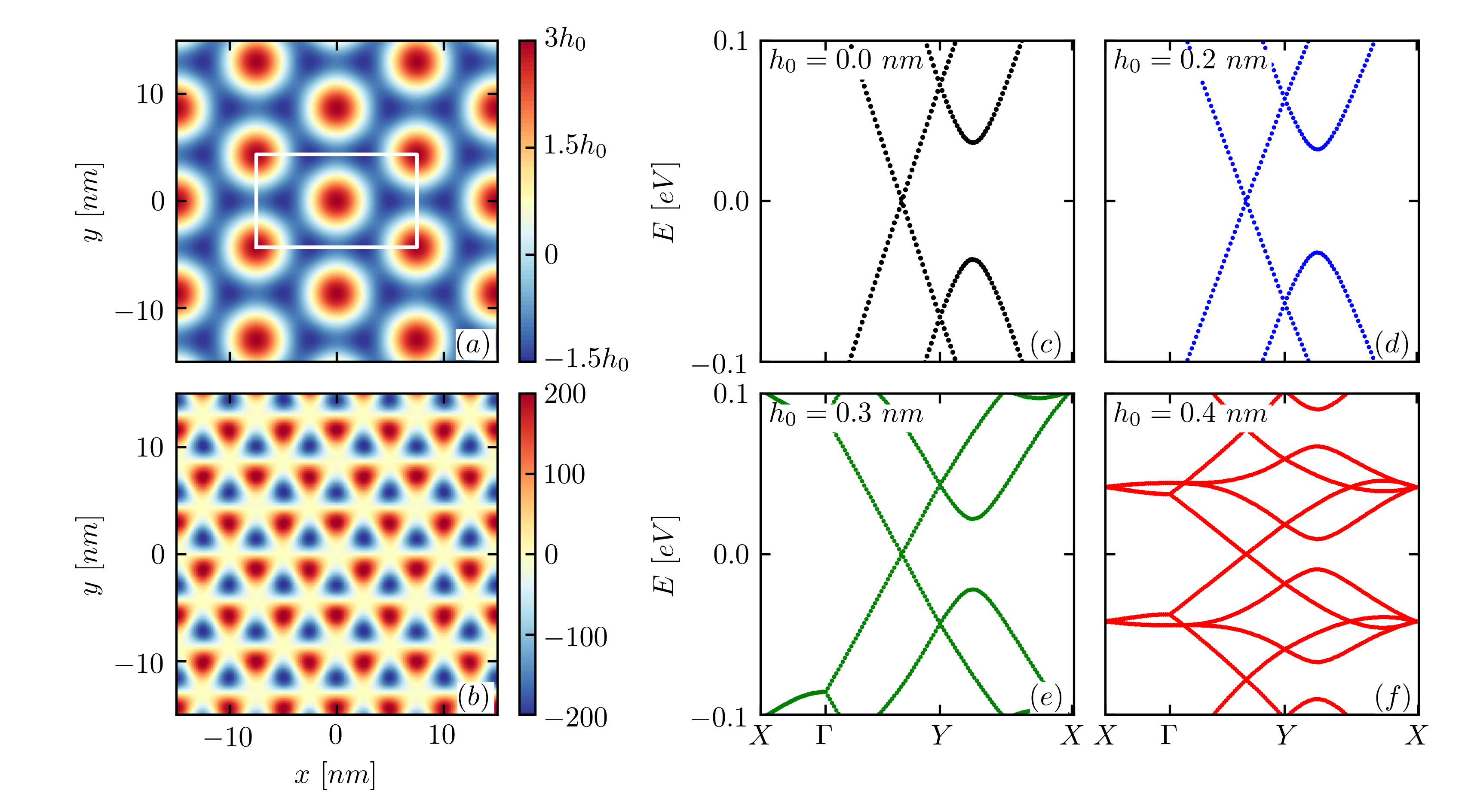}
\caption{(a) The out-of-plane deformation for the hexagonal buckling mode for $a_B = 7.5$ nm. (b) The pseudo-magnetic field profile for the buckling mode from (a) using $h_0 = 0.2$ nm. The units are in Tesla. (c-f) The band structure of the unit cell from (a) for different values of $h_0$ given in the inset.}
\label{fig_4}
\end{center}
\end{figure*}
\section{Hexagonal and herringbone buckling modes}
\label{sec::3}
Next, we turn to the hexagonal buckling mode. The out-of-plane deformation field is given by
\begin{equation}
\label{e_hex_1}
z(x, y) = h_0 \left[ \cos\left(K x\right) + 2 \cos\left(\frac{1}{2}K x\right) \cos\left(\frac{\sqrt{3}}{2} K y\right) \right], \\
\end{equation}
where $h_0$ is the amplitude of the deformation in the $z$-direction and $K = 2\pi/a_B$ is the wave number. This in-plane deformation field was introduced in Ref. \onlinecite{r_bu_mo1}. The previous equation implies that the characteristic wave numbers are $K/2$ in the $x-$direction and $\sqrt{3}/2K$ in the $y-$direction. Hence, the unit cell is a rectangle with length $2a_B$ in the $x-$direction and $2a_B/\sqrt{3}$ in the $y-$direction, as shown in Fig. \ref{fig_4}(a) by the white rectangle. To ensure commensurability with the graphene's unit cell, the unit vectors of the PMF unit cell are given by $\mathbf{a_1} = n_1 \mathbf{a_1^{gr}}$ and $\mathbf{a_2} = n_2 \mathbf{a_2^{gr}}$, where $n_{1, 2}$ are integers and $\mathbf{a_1^{gr}} = (\sqrt{3} a_0, 0)$ and $\mathbf{a_2^{gr}} = (0, 3 a_0)$ are the unit vectors of the 4-atom unit cell of graphene. Notice that the out-of-plane deformation generated by this mode resembles the PMF of the previous buckling mode shown in Fig. \ref{fig_1}(a). The induced pseudo-magnetic field generated by this buckling mode is shown in Fig. \ref{fig_4}(b) and consists of alternating regions of positive and negative PMF of the same strength and with zero field in the center of the bump, similar to the ones obtained for Gaussian bumps \cite{r_gb_1, r_gauss_2} and bubbles \cite{r_bubble_1}. Using Eqs. \eqref{ch8e23} and \eqref{ch8e3} together with the expressions for the deformation field given in Ref. \onlinecite{r_bu_mo1} we obtain analytical expression for the pseudo-magnetic field given by

\begin{equation}
\label{e_pmf_hex}
\begin{split}
&B_{PMF}(x, y) = \frac{A_0}{64} K^3 h_0^2 \left(- 24 \cos\left[\sqrt{3} K y\right] \sin[Kx] - \right.\\
 &\left. 6 (P_1 + P_2 +2 P_1 \cos[Kx])\cos\left[\frac{\sqrt{3}}{2}K y\right] \sin\left[\frac{K x}{2}\right]+ \right.\\
&\left. 2 \sqrt{3} \left(3 P_2 \cos\left[\frac{K x}{2}\right]+(-32+9 P_1) \cos\left[\frac{3 K
x}{2}\right]+ \right. \right. \\
&\left. \left. 8 (4+3 \cos[K x]) \cos\left[\frac{\sqrt{3}}{2} K y\right]\right) \sin\left[\frac{\sqrt{3}}{2} K y\right]\right)
\end{split}
\end{equation}
with $P_1 = (3 - \nu)$, $P_2 = (1 - 3\nu)$, with $\nu$ being the Poisson's ratio of graphene. We should add that this expression is only valid in case of small strains, i.e. when it is justified to approximate Eq. \eqref{e_vecpot} with Eq. \eqref{ch8e23} and it is derived for plotting purposes only. Note that in our tight-binding calculations we always use the full expression for the vector potential.
The band structure calculated for a few values of $h_0$ is given in Figs. \ref{fig_4}(c-f). Notice that the Dirac cone is still present. Increasing the out-of-plane deformation leads to a decrease of the slopes of linear bands in graphene and the formation of semi-flat bands for sufficiently large $h_0$. However, a big difference with the previous case of triangular PMF is that the spectrum stays continuous, i.e. mini-gaps do not open even for a very large straining above 20$\%$. Therefore, in the following we will focus on the DOS plots instead of bandwidth.
\begin{figure*}[htbp]
\begin{center}
\includegraphics[width=17.5cm]{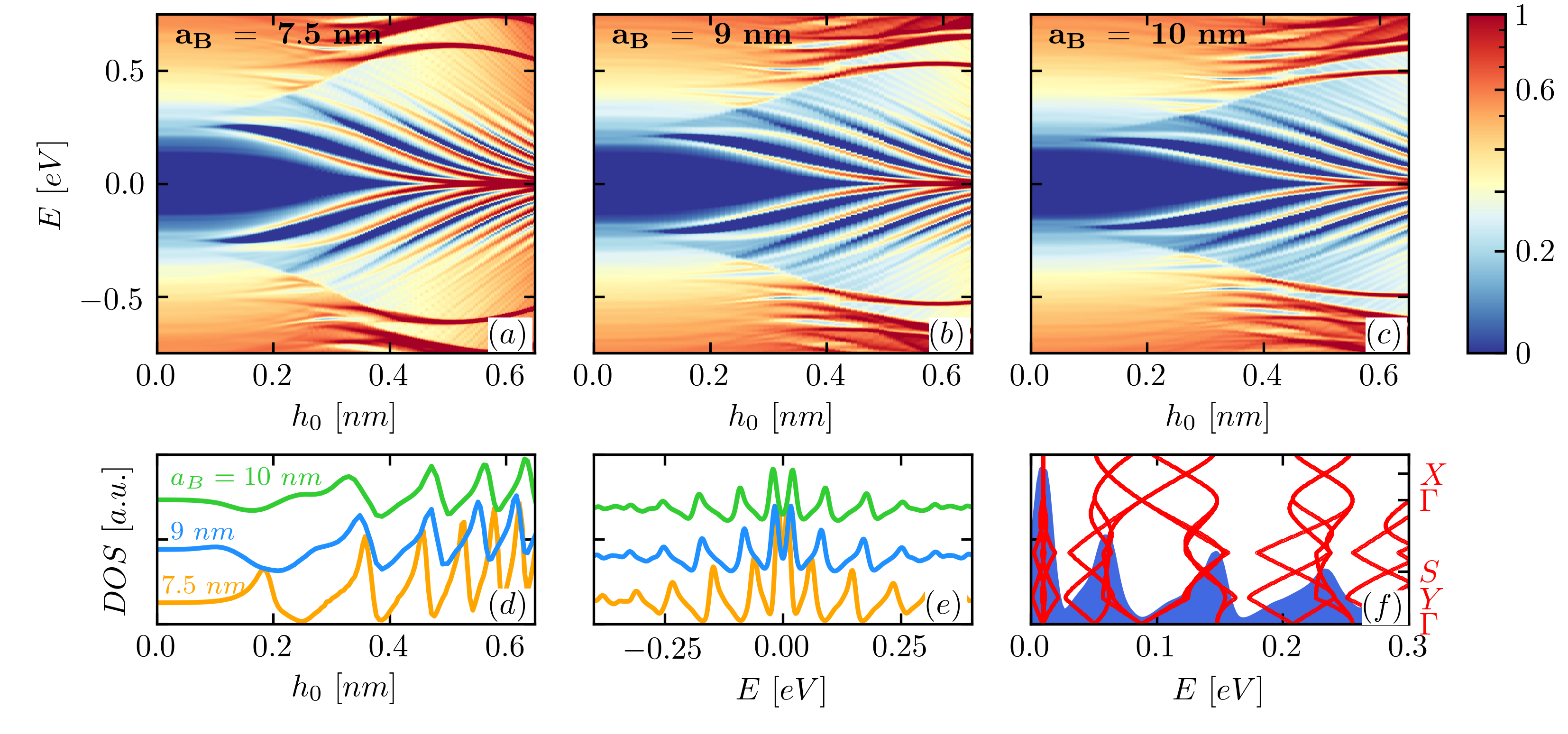}
\caption{Contour plot of the DOS versus the height of the bump and energy for a hexagonal buckling mode with period (a) $a_B = 7.5$ nm, (b) $a_B = 9$ nm, and (c) $a_B = 10$ nm. Cuts of the DOS at (d) constant energy of $E=0.2$ eV and (e) constant amplitude of the deformation, $h_0 = 0.5$ nm. Curves are shifted for better visibility. (f) A cut of the DOS (blue curve) together with the band structure (red curves) for $h_0 = 0.5$ nm and $a_B = 7.5$ nm.} 
\label{fig_5}
\end{center}
\end{figure*}

The DOS versus the amplitude of the out-of-plane deformation and the energy is shown in Figs. \ref{fig_5}(a-c) for three values of $a_B=$ 7.5, 9, and 10 nm, respectively (have in mind that the actual periods are twice these values in the $x$-direction). We see that the DOS exhibits peaks that decrease towards the zero energy with increase of $h_0$ (analogous to increase $B_0$). These levels are better seen in Figs. \ref{fig_5}(d) and (e) where we plot cuts of the DOS at constant energy of $E = 0.2$ eV and constant $h_0=0.5$ nm, respectively. The peaks are less pronounced, less regular, and less separated in comparison to Figs. \ref{fig_3}(d-f) for the case of the triangular PMF. The reason is that the sub-bands are not as flat, as shown in Fig. \ref{fig_5}(f). One can see that the lowest two bands are rather flat which results in narrow, large peaks in the DOS, however, as we move away from the Dirac point the mini-band dispersion becomes more prominent and peaks in the DOS broaden significantly. This can be explain by the fact that the pseudo-magnetic field in this configuration cannot efficiently localize electrons. By comparing this PMF configuration with the previous one (shown in Figs. \ref{fig_1}(a) and \ref{fig_3a}(a)), one can notice that in this case non-zero PMF regions are separated by relatively large regions with zero field. This is sufficient to scatter electrons but not sufficient enough to localize them.
\begin{figure*}[thbp]
\begin{center}
\includegraphics[width=17.5cm]{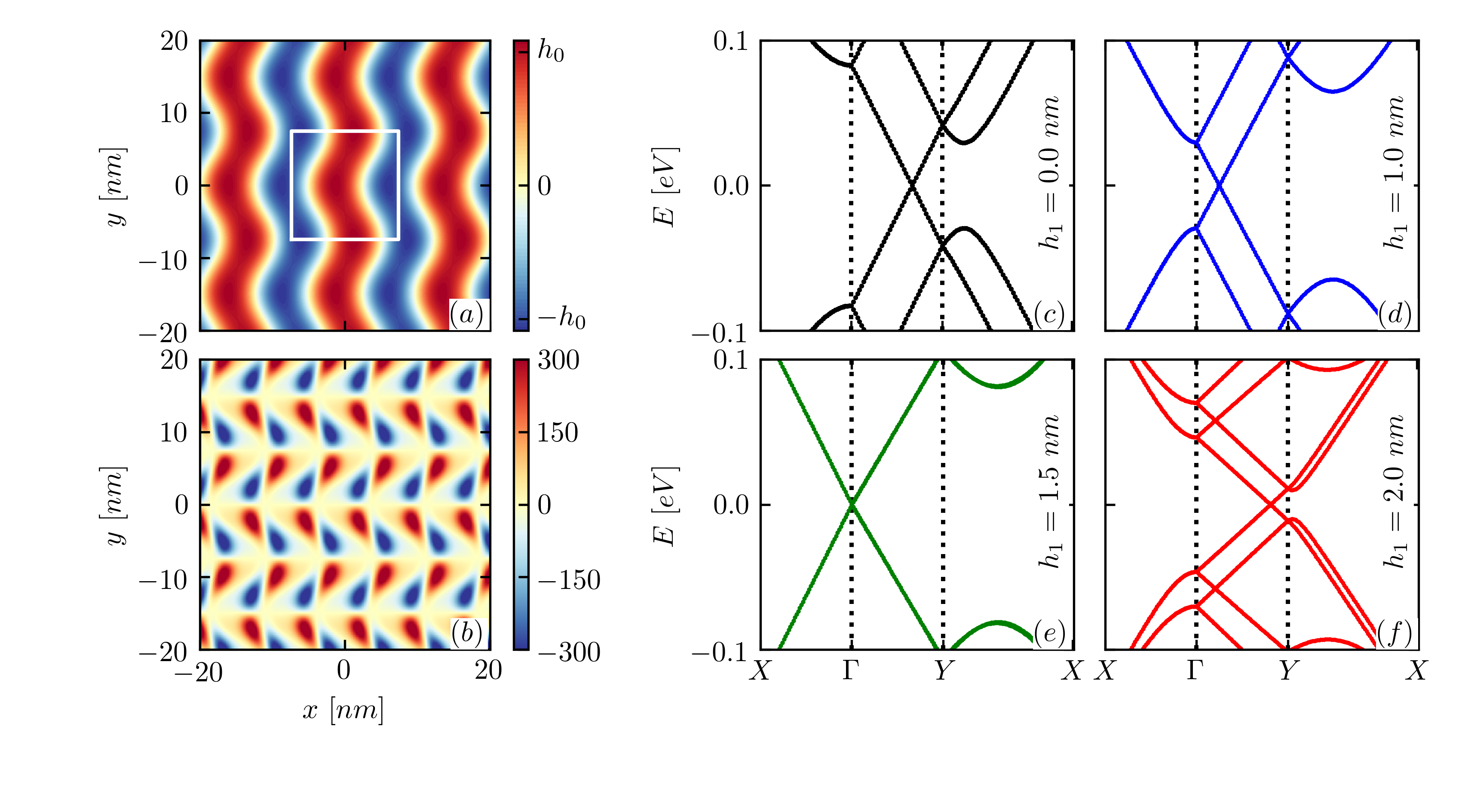}
\caption{(a) Out-of-plane deformation for the herringbone buckling mode with $a_B=15$ nm and $h_2 = 1$ nm. White rectangle shows the unit cell of this mode. (b) Pseudo-magnetic field (in units of Tesla) profile for the buckling mode from (a) using $h_1 = 2$ nm. (c) Band structure for a unit cell from (a) using few values of $h_1$ given in the inset.}
\label{fig_6}
\end{center}
\end{figure*}

Finally, we examine the herringbone buckling mode. This mode is found to be the lowest energy buckling configuration in the case of large biaxial strains \cite{r_bu_mo1}. The out-of-plane displacement, shown in Fig. \ref{fig_6}(a), is given by
\begin{equation}
\label{e_herr_01}
z(x, y)  = h_1 \left[ \cos\left(K_x x \right) - K_x h_2\sin\left( K_x x\right) \cos\left( K_y y\right)  \right],
\end{equation}
where $h_1$ is the amplitude of the out-of-plane deformation, $K_x = 2\pi/a_x$ and $K_y = 2\pi/a_y$ are the wave numbers in the $x-$ and $y-$direction, and $h_2$ defines the breadth of the zigzag pattern, e.g. for $h_2 = 0$ we have a simple $1-$dimensional mode and as we increase $h_2$, the maxima in $z$ (or the corners of the zigzag pattern) move away from $x=0$. The unit cell is shown by a white rectangle in Fig. \ref{fig_6}(a). We note that the unit cell of the PMF is chosen such that the unit vectors are commensurate with an extended 4-atom rectangular unit cell of graphene. The pseudo-magnetic field, shown in Fig. \ref{fig_6}(b), reveals similar positive and negative patterns as in the case of the hexagonal buckling mode with additional distortion in the $x-$direction. Note that in this case the PMF depends on four different variables ($a_x$, $a_y$, $h_1$, and $h_2$). Hence, to reduce the parameter space, we restrict ourselves to $a_x = a_y = a_B$, i.e. superlattice vectors have the same length in both directions and the unit cell is a square. The band structure resulting from the deformation shown in Fig. \ref{fig_6}(a) using few values of $h_1$ is shown in Figs. \ref{fig_6}(c-f). By increasing the strain in the unit cell one squeezes the bands and creates energy windows with high number of available states but no gap opening is found at the Dirac point. Notice that also in this case the Dirac cone survives. No clear, narrow energy bands appears.
\begin{figure*}[htb]
\begin{center}
\includegraphics[width=17.5cm]{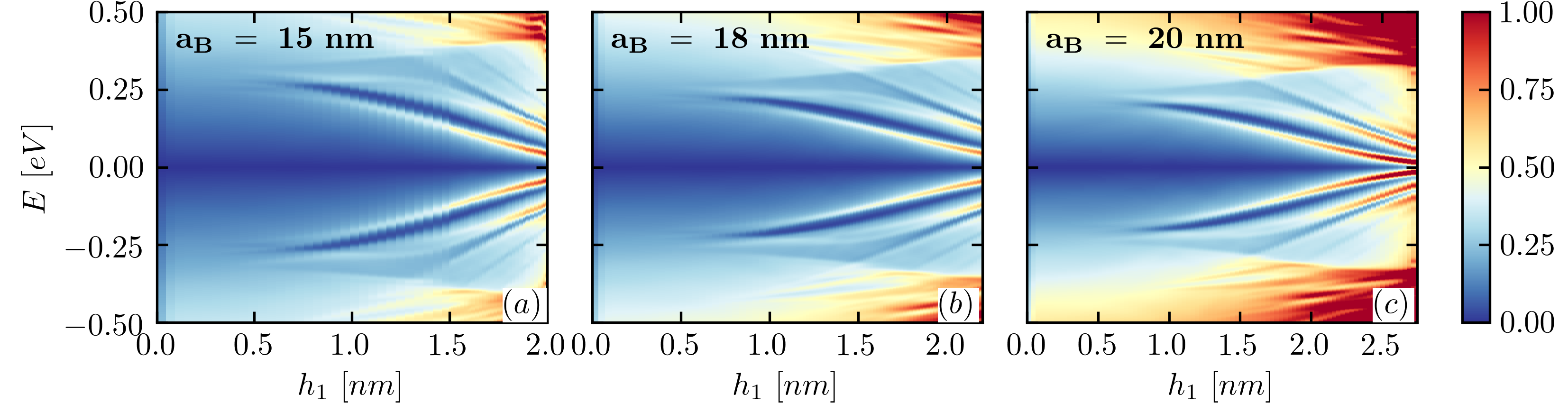}
\caption{Contour plot of the DOS versus the height of the bump and energy for the herringbone buckling mode with (a) $a_B = 15$ nm, (b) $a_B = 18$ nm, (c) $a_B = 20$ nm and fixed $h_2 = a_B/10$.}
\label{fig_7}
\end{center}
\end{figure*}

In Figs. \ref{fig_7}(a-c) we plot the DOS versus the energy and $h_1$. The maximal value of $h_1$ in each plot is chosen such that the maximal strain stays below 25 $\%$ which is the breaking limit for graphene. Figures show that this buckling mode does not favour the appearance of flat bands. High DOS states are observed only at large values of $h_1$. Furthermore, no clear zeroth LL develops in this buckling mode even when strain reaches 20 $\%$. One can understand this result as a consequence of the fact that this is the lowest energy configuration of all buckling modes, having minimal total energy (sum of membrane, bending, and cohesive energy), in the case of large strain. In other words, since all three components of the total energy are directly related to the elements of the strain tensor \cite{r_bu_mo1}, the overall strain induced by this buckling mode is the lowest compared to other buckling modes. To test this, we calculate the average strain within the hexagonal and the herringbone unit cell of approximately the same area induced when the maximal out-of-plane deformation reaches 2 nm. The average strain in the hexagonal unit cell is double the value in the herringbone unit cell across the wide range of superlattice periods and buckling amplitudes. Furthermore, if we compare the pseudo-magnetic field of the two cases the difference is even more pronounced. An order of magnitude higher PMF (maximal value and absolute average value) is achieved in the case of hexagonal buckling.
\section{Spatial LDOS and current}
\label{sec::4}
\begin{figure*}[htb]
\begin{center}
\includegraphics[width=17.5cm]{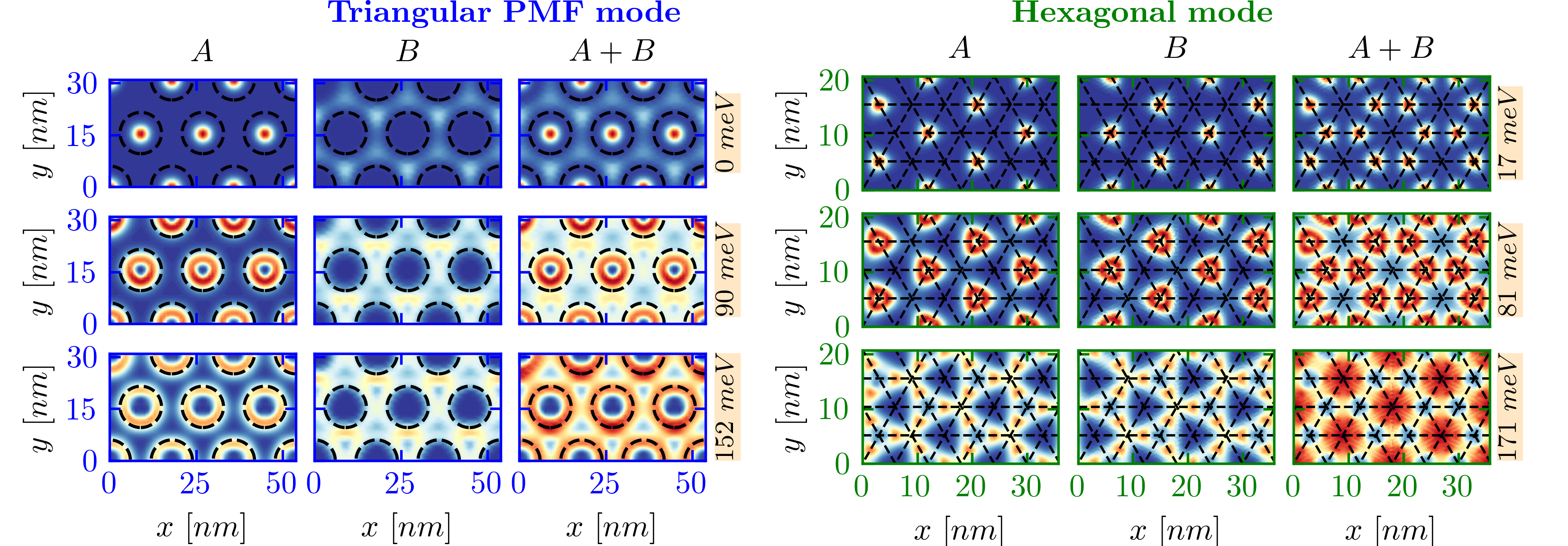}
\caption{Spatial LDOS for three lowest flat bands for the triangular PMF mode (left panel) and the hexagonal buckling mode (right panel). The maps are calculated using $a_B = 18$ nm and $B_0 = 100$ T for the triangular PMF mode and $a_B= 9$ nm and $h_0 = 0.5$ nm for the hexagonal buckling mode at energies shown to the right of the figures. LDOS is shown for both sublattices separately as well as total LDOS. Dashed black curves indicate $B(\mathbf{r})=0$ cuts.}
\label{fig_8}
\end{center}
\end{figure*}
In this section, we present the spatial local density of states maps that give us a visual insight into the distribution of states and their dependence on buckling parameters. Left panel in Fig. \ref{fig_8} shows spatial LDOS maps for energies corresponding to the three lowest flat bands shown in Figs. \ref{fig_3}(b) and (f) (orange curve) calculated for $B_0=100$ T and integrated over the whole Brillouin zone. Different columns show the spatial LDOS on two sublattices separately (left and middle column) and total LDOS (right column). Notice that the PMF breaks the sublattice symmetry \cite{r_gb_3}. In the case of the triangular PMF (shown in the left panel), the sign of the PMF varies inside the unit cell with the positive field inside the regions marked by dashed circles and negative field elsewhere. This will have a profound influence on the density of states. The states inside the circles are localized on the A sublattice while in the regions with negative PMF, the states are localized on the B sublattice. Hence, states on opposite sublattices are spatially separated. Similar conclusion can be drawn from the right panel as well. One interesting feature is that the regions with the highest LDOS are localized around $B = 0$ cuts. The reason is that the strong fields scatter electrons away from these regions. The exceptions are the Landau levels, corresponding to closed orbits, which are localized inside regions with the highest field. This is, as we discussed, due to the fact that in the case of spatially varying magnetic fields, the Landau levels start developing at regions where the magnitude of the field is highest, at fields for which the extent of the LL wave-function is smaller or at most of the order of the size of these regions.
\begin{figure*}[tbhp]
\begin{center}
\includegraphics[width=17.5cm]{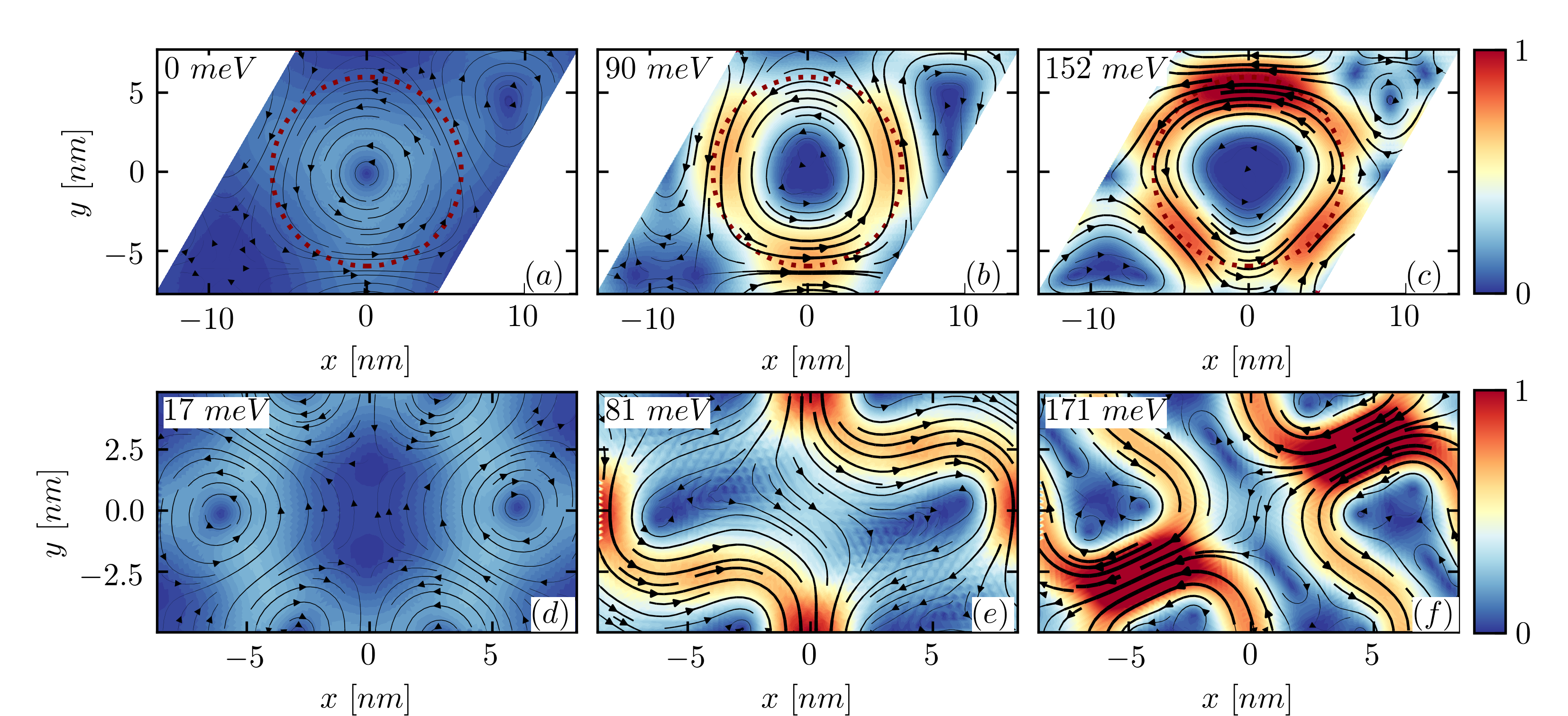}
\caption{Inter-atomic current (shown in arbitrary units) within the PMF unit cell for three lowest flat bands (energy values are given in each plot) at the $\mathbf{K}$ point for the triangular PMF mode (a-c) and the hexagonal buckling mode (d-f). Current vectors are shown by black lines and a color contour plot gives the intensity of the current. Dotted maroon circles in (a-c) show zero-PMF cuts.}
\label{fig_9}
\end{center}
\end{figure*}

As already shown by the spatial LDOS maps, the states induced by the periodically buckled lattices are quasi-localized and percolate throughout the whole unit cell. In other words, the strain superlattice creates conducting paths for the charge carriers. To prove this, in Fig. \ref{fig_9} we plot a k-dependent inter-atomic current calculated using
\begin{equation}
\label{e_iac}
\mathbf{j}_{\mathbf{k}, i}(E) = \frac{4e}{h} \sum_{j=1}^3 \operatorname{Im} \left[ \Psi_{\mathbf{k}, i}^*(E) H_{ij} \Psi_{\mathbf{k}, j}(E) \right]\mathbf{e}_{ij},
\end{equation}
where $\mathbf{j}_{\mathbf{k}, i}(E)$ is the current at position $i$ calculated as a sum of currents flowing between atom $i$ and its three nearest neighbours at energy $E$ due to an electron in state $\mathbf{k}$, $e$ is the elementary charge, $h$ is the Planck constant, $ \Psi_{\mathbf{k}, i}$ is the wave function of the $\mathbf{k}-$state at atom $i$, $H_{ij}$ is the Hamiltonian element between atoms $i$ and $j$, and $\mathbf{e}_{ij}$ is the unit vector in the direction of the bond between atoms $i$ and $j$. In Fig. \ref{fig_9} we plot the current due to electrons of the $\mathbf{K}$ valley with wavevector $\mathbf{k} = \mathbf{K}$ and the same values of the energy as in Fig. \ref{fig_8}. Note that due to the flatness of the bands, the current in other k-points has a similar profile as in $\textbf{K}$. However, due to the sublattice symmetry breaking, the valley degeneracy is lifted and, hence, different sub-bands that appear in Fig. \ref{fig_1}(c) have opposite symmetries of the wave-functions (blue and red curves) in the two valleys. Consequently, the current direction will be opposite in the two valleys. This leads to zero total current in the band, as expected since time-reversal symmetry is not broken by the strain fields. This leads to zero Hall resistance in the band (however, the longitudinal resistance will be non-zero).
% %
% \begin{figure*}[tbhp]
% \begin{center}
% \includegraphics[width=15cm]{figs/fig_10_a.png}
% \caption{LDOS at a point in the center of the triangular PMF unit cell with $a_B = 18$ nm. Top row shows the LDOS versus the amplitude of the PMF and energy with constant real field: (a) $B_R = 10$ T and (b) $B_R = 20$ T. Bottom row shows LDOS versus the real field and energy with constant amplitude of the PMF: (c) $B_0 = 50$ T and (d) $B_0 = 100$ T. The color bar is in arbitrary units.}
% \label{fig_10}
% \end{center}
% \end{figure*}
%
\begin{figure*}[htbp]
\begin{center}
\includegraphics[width=7.5cm]{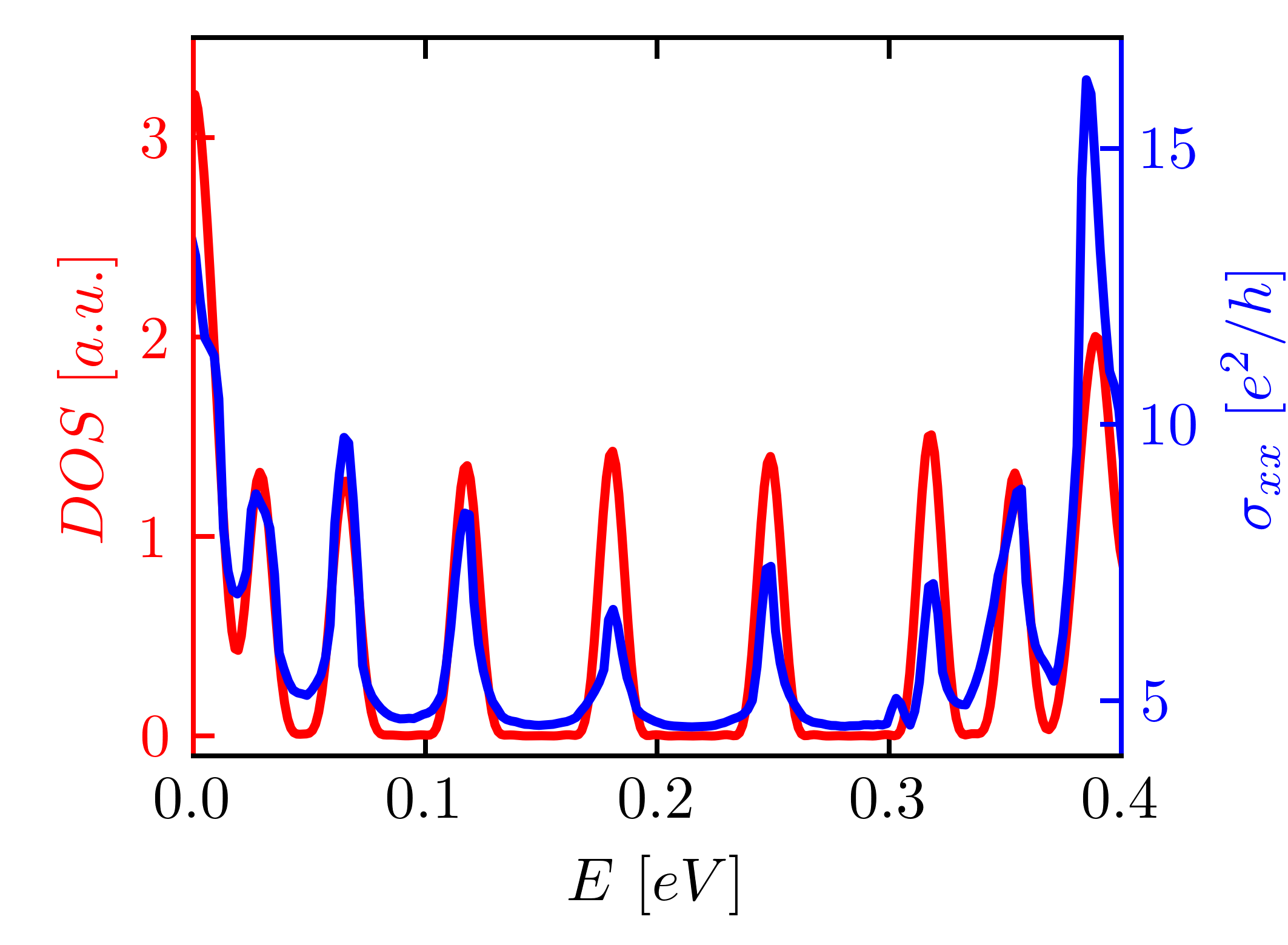}
\caption{The longitudinal conductance, $\sigma_{xx}$, (blue curve) and the density of states (red curve) for a periodically strained graphene using triangular PMF profile with $a_B = 18$ nm and $B_0 = 150$ T.}
\label{fig_10a}
\end{center}
\end{figure*}

However, current carriers are not localized like in the case of e.g. Landau levels where the carriers in the bulk circle around their equilibrium positions and only the electrons that skip along the edges of the sample contribute to the overall conductance. A similar scenario can be seen in Figs. \ref{fig_9}(a) and (d) where the current is negligible even though spatial LDOS plots show high density of states at these energies (Fig. \ref{fig_8} top row). These electrons are confined by the strong strain fields and cannot propagate. As we move to higher bands (rest of plots in Fig. \ref{fig_9}), the situation changes and strong currents can be observed flowing between regions with high PMF values. This creates percolating paths along which electrons can propagate in both directions. To confirm that the flat bands do not correspond to localized states we calculate the longitudinal conductance, $\sigma_{xx}$, using the Kubo-Bastin \cite{cKB} formula at zero temperature and compared it against the DOS. Fig. \ref{fig_10a} shows these two quantities for a triangular PMF unit cell with $a_B = 18$ nm and $B_0 = 150$ T for an energy broadening of 5 meV. The positions of prominent peaks in the conductance agree perfectly with the peaks in the DOS and undoubtedly confirm the influence of flat bands on electronic transport. Consequently, transport measurements are able to detect flat band states. 
\begin{figure*}[htbp]
\begin{center}
\includegraphics[width=16cm]{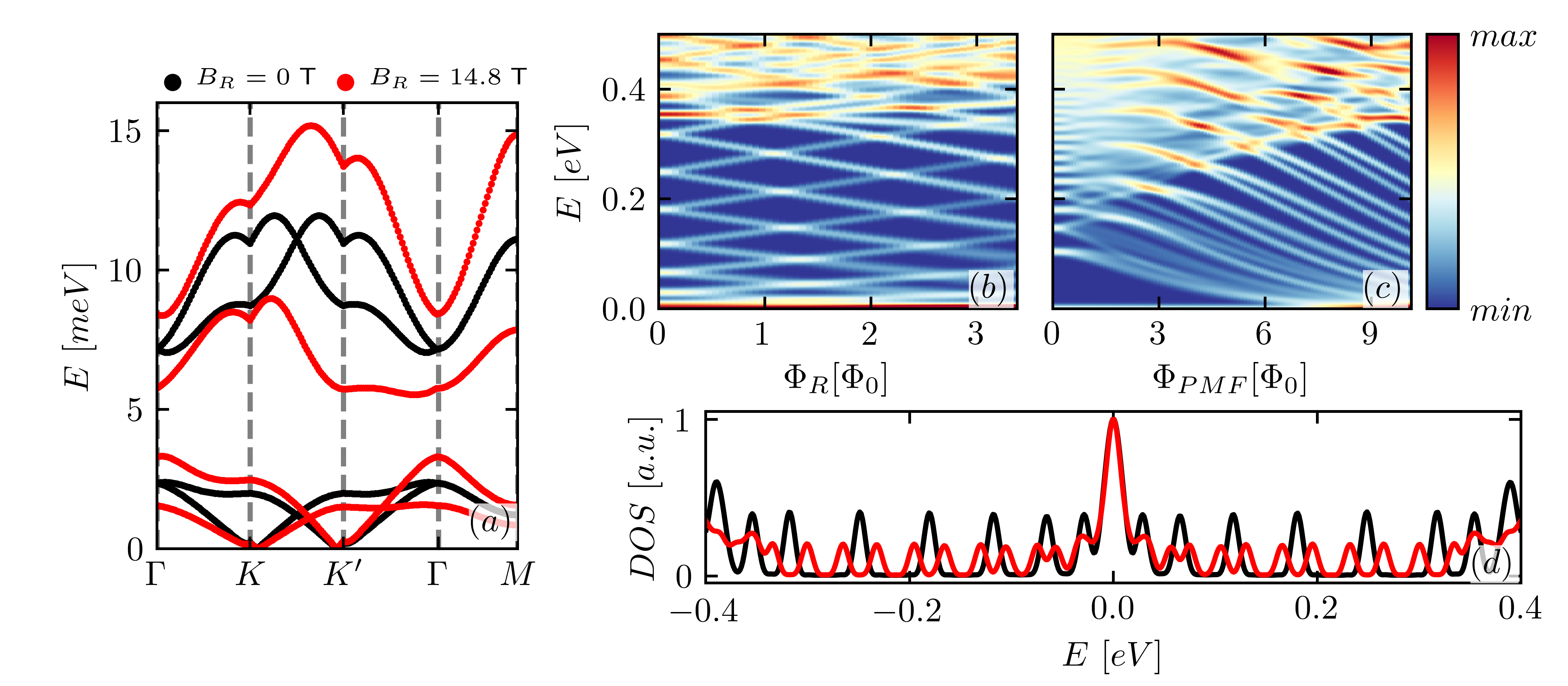}
\caption{(a) The band structure of the triangular PMF unit cell with $a_B = 18$ nm and $B_0 = 150$ T in the presence of a uniform real magnetic field $B_R$ shown in the inset. (b) The contour plot of the LDOS in the dark region versus the flux generated by real magnetic field flux $\Phi_R$ (in units of magnetic flux quantum) and energy using $a_B = 18$ nm and $B_0 = 150$ T. (c) The contour plot of the LDOS versus the flux of the pseudo-magnetic field and energy using $a_B = 18$ nm and $B_R = 10$ T. (d) Total DOS versus energy for a PMF unit cell from (a) in the absence of real magnetic field (black curve) and in the presence of small magnetic field $B_R = 5.8$ T (red curve) using broadening of 5 meV.}
\label{fig_12}
\end{center}
\end{figure*}
 \section{Valley polarized states}
\label{sec::4a}
As already mentioned, the presence of the PMF does not lift the valley degeneracy of the system. Fig. \ref{fig_1}(c) showed that the spectrum of a periodically strained system consists of two sets of bands (red and blue) belonging to different valleys of the system. PMFs in these valleys have equal amplitude but opposite direction. By applying the perpendicular real magnetic field, $B_R$, one can lift this degeneracy since the effective field in two valleys, $B_{eff} = B_R \pm B_{PMF}$, will differ i.e. in one valley the real and pseudo-field will be in the same direction while in the other they will be in opposite directions. This is shown in Fig. \ref{fig_12}(a) where the low energy spectrum of the PMF unit cell is shown in the absence (black curves) and presence (red curves) of the real magnetic field. The amplitude of the magnetic field is chosen such that the magnetic flux through the buckling unit cell corresponds to one flux quantum, i.e. $B_RS = \Phi_0$, where $S$ is the area of the PMF unit cell and $\Phi_0$ is the magnetic flux quantum. In case of $a_B = 18$ nm, we find that the value of the real magnetic field that generates this flux is $B_R = 14.8$ T. When comparing two spectra it is obvious that one set of bands moves downwards while the other set moves upwards in energy. This results in valley polarized bands which are of great importance for valleytronics \cite{r_val_01, r_val_02, r_val_03, r_val_04}. In Fig. \ref{fig_12}(b) we plot the local density of states in the dark region versus the real field, $B_R$, and energy, using $a_B = 18$ nm and $B_0 = 150$ T (corresponding to $\Phi_{PMF} = 10.1\Phi_0$). We see that in the case of $B_R = 0$, LDOS shows series of well-defined equidistant peaks - confinement states. As the real magnetic field is introduced, these peaks split into 2 states from which one moves downwards while the other moves upwards in energy with the increase of $B_R$, in agreement with the behaviour of the bands from Fig. \ref{fig_12}(a). Similar behavior is observed in Fig. \ref{fig_12}(c) where the contour plot of LDOS is shown versus the amplitude of the PMF and energy using constant value of the real field, $B_R = 10$ T. In this case, we start with the Landau levels (for $B_0 = 0$ T) which turn into confinement states as the amplitude of the PMF is increased, similarly as in Fig. \ref{fig_3}(a-c), but with a major difference that each confinement state is spit into two states by the magnetic field. The splitting of the confinement states in the presence of the real magnetic field is not a local effect, which we confirm by calculating total density of states. Fig. \ref{fig_12}(d) shows DOS for a PMF unit cell in the absence (black curve) and presence of a small real magnetic field (red curve). One can immediately notice the splitting of the peaks in the DOS due to the interplay between the real and the pseudo-magnetic field. This effect was recently experimentally confirmed in Refs. \onlinecite{r_val_15, r_val_16}.
\section{Conclusions}
\label{sec::5}
In this paper we determined the conditions for the appearance of flat bands in periodically buckled graphene systems. We considered three different buckling modes. For the triangular PMF mode we showed that the linear low energy spectrum (i.e. the Dirac cone) is transformed into a series of mini-bands by the periodic vector potential. We examined the lowest three bands and found that a flux of $6.5\Phi_0$ through the PMF unit cell is needed to flatten the band in order that electron-electron interactions become important. On the other hand, the two higher bands are much flatter and fluxes of only $4.4\Phi_0$ and $1.7\Phi_0$, respectively, are needed to achieve regimes where interactions could become stronger than the kinetic energy of the electrons. Plots of the DOS versus the magnitude of the PMF and energy revealed the existence of states in-between Landau levels. These mimic the behaviour of electronic states of graphene quantum dots in a real magnetic field. We showed that these states correspond to the flat bands and the connection with the quantum dot system is due to the electron confinement resulting from the strong non-uniform strain.

The hexagonal and the herringbone buckling modes showed rather different effects on the electron properties. Namely, the DOS versus the amplitude of the out-of-plane deformation and the energy revealed similar electron states which demonstrates that their appearance is not due to a specific choice of the vector potential but can be found in various periodic buckling configurations. However, unlike the first PMF configuration, these two buckling modes proved to be less favourable systems for the realization of flat bands. In the case of the hexagonal buckling mode, only a few lowest energy bands had the tendency to become flat while in the case of the herringbone mode, flatness of the bands appeared only at very large non-uniform periodic strain. We compared the values of the PMF generated for equal value of the out-of-plane deformation and found that the herringbone mode showed an order of magnitude smaller PMF fields as compared to the hexagonal mode. This supports the idea that strong pseudo-magnetic field-induced confinement is needed for the creation of flat bands.

We calculated the spatial LDOS maps of few lowest flat bands and the electron current corresponding to those flat band states. We confirmed that there exist electronic flat band states that percolate throughout the whole system. This behaviour is contrary to Landau level states which are highly localized within the unit cell. This was also supported by the single state current plots where we showed that strong currents flow through the unit cell - resulting in percolating paths throughout the system. Non-localized nature of these states was confirmed by conductance calculations which showed prominent peaks at energies that correspond to flat band states. Upon applying real magnetic field, the confinement states split due to the lifting of the valley degeneracy. This results in valley-polarized flat bands which are of great interest for valleytronics. 
\section{Acknowledgment}
One of us (SPM) is supported by the Flemish Science Foundation (FWO). We thank E. Y. Andrei, Y. Jiang, and J. Mao for fruitful discussions.
\appendix
\section{Gauge field for the triangular PMF mode}
\label{sec::s1}
\begin{figure}[htbp]
\centering
\includegraphics[width=8.5cm]{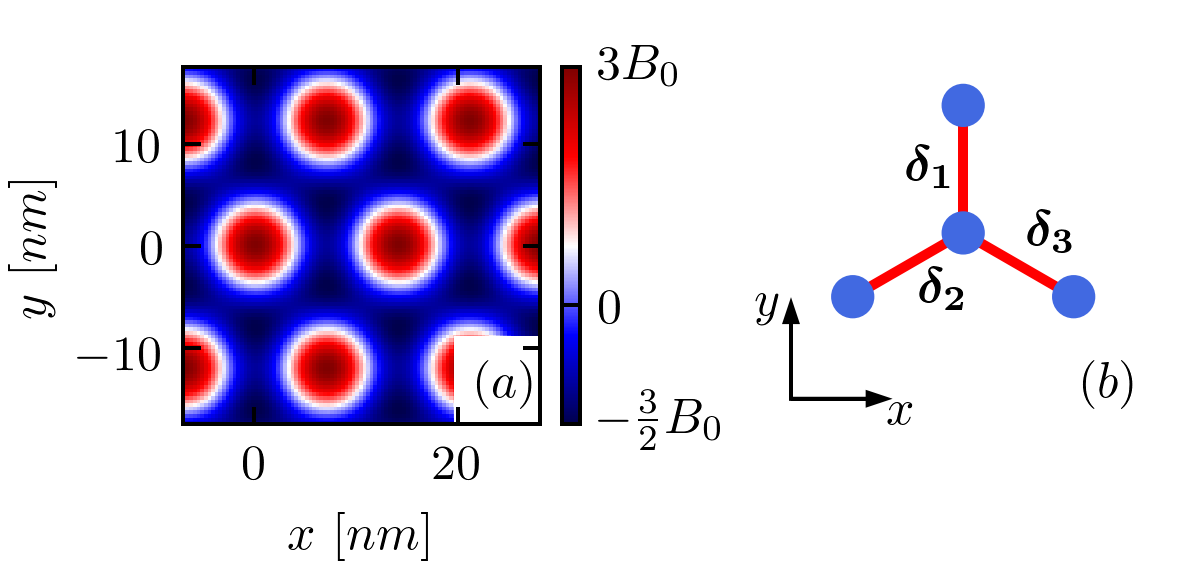}
\caption{(a) Profile of the pseudo-magnetic field given by Eq. (1) of the main manuscript with $a_B = 15$ nm. (b)  Graphene lattice with nearest neighbours vectors $\boldsymbol{\delta_1}$, $\boldsymbol{\delta_2}$, and $\boldsymbol{\delta_3}$.}
\label{fig_guc}
\end{figure}

The effect of strain is included in the tight-binding Hamiltonian through the modulation of the hopping energy. This is given by
\begin{equation}
	\label{ee1}
		t_{ij} = t_0 e^{-\beta(d_{ij}/a_{cc}-1)},	
\end{equation}
where $d_{ij}$ is the strained bond length defined by the strain tensor $\overline{\boldsymbol{\varepsilon}}$ as
\begin{equation}
	d_{ij} = (\mathbf{I} + \overline{\boldsymbol{\varepsilon}})\boldsymbol{\delta_{ij}},
\end{equation}
where $\boldsymbol{\delta_{ij}}$ is the vector in the direction of the bond between atoms $i$ and $j$ and $\mathbf{I}$ is the unitary matrix. Changes in the hopping energies generates a strain induced vector potential in the system which in the case of hexagonal lattices is given by Eq. \eqref{e_vecpot}. Corresponding pseudo-magnetic field is calculated using
\begin{equation}
\label{e_pmf}
\mathbf{B} = \pmb{\bigtriangledown} \times \mathbf{A} = \left[\partial_x A_y - \partial_y A_x\right]\mathbf{e_z}.
\end{equation}
Due to gauge invariance we may choose $A_y = 0$. Hence, our vector potential is then given by $A_x = \int B(x, y) dy$ where $B(x, y)$, shown in Fig. \ref{fig_guc}(a), is given by Eq. \eqref{e_tpmf_1} of the main manuscript. This leads to
\begin{equation}
\label{e_ax}
\begin{split}
	A_x = B_0 \frac{a_B}{2\pi}\left[
						      \frac{1}{K_{1y}}\sin(\boldsymbol{K_1}\boldsymbol{r})+
						      \frac{1}{K_{2y}}\sin(\boldsymbol{K_2}\boldsymbol{r})+ \right. \\
						      \left. + \frac{1}{K_{3y}}\sin(\boldsymbol{K_3}\boldsymbol{r}) \right].
\end{split}
\end{equation}

Substituting  $t_{ij} = t_0(1 + \delta t_{ij})$ and expanding Eq. \eqref{e_vecpot} up to first order, we obtain the following expression for the vector potential \cite{ref_01}
	\begin{equation}
	(A_x, A_y) = -\frac{1}{2ev_F}\left(2\delta t_1 - \delta t_2 - \delta t_3, \frac{1}{\sqrt{3}}(\delta t_2 - \delta t_3)\right),
	\end{equation}	 
where $\delta t_1$, $\delta t_2$, and $\delta t_3$ are the strain modulations of hopping energies along the directions of graphene's nearest neighbours $\boldsymbol{\delta_1}$, $\boldsymbol{\delta_2}$, and $\boldsymbol{\delta_3}$, as shown in Fig. \ref{fig_guc}(b), and $v_F = 3t_0 a_{cc}/(2\hbar)$ is the Fermi velocity. The choice of the gauge ($A_y=0$) results into $\delta t_2 = \delta t_3 = \delta t$. We choose $\delta t_1 = - \delta t$ and, finally, the strain modified hopping energies are given by 
	\begin{equation}
	\begin{split}
		t_1 &= t_0 \left(1 - \frac{3A_x \pi  a_{cc}}{2\phi_0}\right) \\
		t_2 = t_3 &= t_0 \left(1 + \frac{3A_x \pi  a_{cc}}{2\phi_0}\right), \\
		\end{split}
	\end{equation}	  
where $\phi_0 = h/e$ is the magnetic flux quantum.
\pagebreak
% \widetext
% \begin{center}
% \begin{huge}
%  \textbf{Appendices}
%  \end{huge}
% \end{center}
% %%%%%%%%%% Merge with supplemental materials %%%%%%%%%%
% %%%%%%%%%% Prefix a "S" to all equations, figures, tables and reset the counter %%%%%%%%%%
% \setcounter{equation}{0}
% \setcounter{figure}{0}
% \setcounter{table}{0}
% \setcounter{section}{0}
% \setcounter{page}{1}
% \makeatletter
% \renewcommand{\theequation}{A\arabic{equation}}
% \renewcommand{\thefigure}{A\arabic{figure}}
% \renewcommand{\thesection}{A\arabic{section}}
   
%
%
%
\end{document}